\documentclass[useAMS,usenatbib]{mn2e}

\usepackage{dsfont}
\usepackage{amsmath}
\usepackage{amssymb}
\usepackage{graphicx}
\usepackage{verbatim}
\usepackage{natbib}
\usepackage{eucal}
\usepackage{calligra}

\usepackage{amsfonts}
\usepackage[usenames,dvipsnames]{color}
\usepackage[normalem]{ulem}
\usepackage[T1]{fontenc}

\DeclareMathAlphabet{\mathscr}{OT1}{pzc}%
                                 {m}{it}
                                 
\newcommand{\mnras}{MNRAS}
\newcommand{\jcap}{JCAP}
\newcommand{\apj}{ApJ}
\newcommand{\apjs}{ApJS}
\newcommand{\aj}{AJ}
\newcommand{\aap}{A\&A}
\newcommand{\prd}{Phys. Rev. D}

\def\apjl{ApJL}

\newcommand{\be}{\begin{equation}}
\newcommand{\ee}{\end{equation}}
\newcommand{\bes}{\begin{equation*}}
\newcommand{\ees}{\end{equation*}}
\newcommand{\bea}{\begin{eqnarray}}
\newcommand{\eea}{\end{eqnarray}}
\newcommand{\beas}{\begin{eqnarray*}}
\newcommand{\eeas}{\end{eqnarray*}}

\newcommand{\mpch}{\;{\rm Mpc}/h}
\newcommand{\mhpc}{M_\odot h / {\rm pc}^2}
\newcommand{\de}{{\rm d}}

\newcommand{\zl}{z_{\rm L}}
\newcommand{\zs}{z_{\rm s}}

\newcommand{\dsop}{\Delta\Sigma^{(+)}_1}
\newcommand{\dsom}{\Delta\Sigma^{(-)}_1}
\newcommand{\dstp}{\Delta\Sigma^{(+)}_2}
\newcommand{\dstm}{\Delta\Sigma^{(-)}_2}

\def\ave#1{\left\langle #1 \right\rangle}

\begin{document}

\title[Halo Ellipticity]
  {Lensing Measurements of the Ellipticity of LRG Dark Matter Halos}
\author[Clampitt and Jain]
  {Joseph~Clampitt$^{1,}$\thanks{E-Mail: clampitt@sas.upenn.edu}, Bhuvnesh~Jain$^{1}$ \\
  $^1$Department of Physics and Astronomy, Center for Particle Cosmology, \\
  University of Pennsylvania,
  209 S. 33rd St., Philadelphia, PA 19104, USA}

\maketitle

\begin{abstract}
Lensing measurements of the shapes of dark matter halos can provide tests of gravity theories and possible dark matter interactions.
We measure the quadrupole weak lensing signal from the elliptical halos of 70,000 SDSS Luminous Red Galaxies.
We use a new estimator that nulls the spherical halo lensing signal, isolating the shear due to anisotropy in the dark matter distribution.
One of the two Cartesian components of our estimator is insensitive to the primary systematic, a spurious alignment of lens and source ellipticities, allowing us to make robust measurements of  halo ellipticity.
Our best-fit value for the ellipticity of the surface mass density is $0.24 \pm 0.06$, which translates to an axis ratio of 0.78.
We rule out the hypothesis of no ellipticity at the $4\sigma$ confidence level, and ellipticity < 0.12 (axis ratio > 0.89) at the $2\sigma$ level. 
We discuss how our  measurements of halo ellipticity are revised to higher values using estimates of the misalignment of mass and light from simulations. 
Finally, we apply the same techniques to a smaller sample of redMaPPer galaxy clusters and obtain a $3\sigma$ measurement of cluster ellipticity.
We discuss how the improved signal to noise properties of our estimator can enable  studies of halo shapes for different galaxy populations with upcoming surveys. 
\end{abstract}

\begin{keywords}
cosmology: observations -- dark matter;
gravitational lensing: weak;
galaxies: haloes
\end{keywords}

\section{Introduction}
\label{sec:intro}

While isotropic dark matter distributions can accurately describe galaxy-galaxy lensing statistics of halos \citep{mhs2005, jsw07,sjm09}, there is a theoretical expectation for individual halos to have significant ellipticity \citep{js02}.
A number of authors have attempted to measure this higher-order signal on galaxy-scale \citep{hyg04,mhb06,phh07,vhs12} and cluster-scale halos \citep{eb09,oto10}.
The cluster measurements, where the intrinsic signal is expected to be much stronger, have been more successful.
Using 4,300 cluster lenses \citet{eb09} obtain a $2.7\sigma$ detection of signal.
Furthermore, using a subset of 18 clusters that are well fit by an elliptical NFW model, \citet{oto10} estimate a $7\sigma$ detection of ellipticity for those specific clusters.

Missing from the literature until now is an attempt to detect and characterize the ellipticity of group-size halos, such as those in which Luminous Red Galaxies (LRGs) reside.
With intermediate size between galaxies and clusters, these LRG halos have larger intrinsic lensing signal than the galaxy samples and are larger in number than the cluster samples.
However, as with the galaxy halos, these have neither enough member galaxies nor enough unstacked lensing signal to make an initial guess for the halo major axis.
We must rely on the ellipticity of the light distribution in order to perform an aligned, stacked lensing measurement.
This approach has both advantages and disadvantages: on the positive side our method has promise in constraining MOdified Newtonian Dynamics \citep{m83} and related modified gravity theories \citep{k15}, which rely on the distribution of baryonic matter alone to produce every observable.
Such theories will have difficulty producing sufficient anisotropic lensing signal at scales much greater than the halo's member galaxies.
On the other hand, this method is susceptible to a possible systematic alignment of lens and source ellipticities \citep{mhb06} and constraints must take account of possible misalignment of the dark matter and light distributions \citep{acd15}.
Nonetheless, we show that our method and resulting lower bounds on halo ellipticity are robust even in the presence of these systematics.

The measurement in this paper is along the lines of our previous work producing lensing detections that go beyond the description of isotropic dark matter halos to measure higher order structures of the cosmic mass distribution.
In \citet{clampitt16} we detected the weak lensing signal of filaments, and in \citet{cj15} we constrained the density profile of cosmic voids.
For all these measurements, filament lensing, void lensing, and now elliptical halo lensing, we have relied heavily on the Sloan Digitial Sky Survey\footnote{http://www.sdss.org} (SDSS) shear catalogs of \citet{sjs2009}.
While new lensing measurements from the Dark Energy Survey\footnote{http://www.darkenergysurvey.org} (DES) are starting to be released \citep{msh15, cvj15, vcj15} (as well as the Kilo Degree Survey\footnote{http://kids.strw.leidenuniv.nl} and the Subaru\footnote{http://www.subarutelescope.org/index.html} HSC survey), there is still more to be gleaned from the previous generation data of SDSS.
Furthermore, the measurement techniques developed in these works will be useful to apply to DES and other cutting edge datasets in the near future.

In \S~2 we describe our model for the shear from elliptical halos as well as potential systematic errors.
In \S~3 we describe our data for lens and source galaxies and also specify our observables.
Then in \S~4 we show the results for the measurement and constraints on halo ellipticity, compare to previous estimators used in the literature, and repeat the measurement with redMaPPer clusters.
In \S~5 we discuss systematic shear tests and the impact of galaxy-halo misalignment.
Finally, in \S~6 we summarize our main conclusions, compare to previous work, and discuss implications for theories of gravity and dark matter.

\section{Model}
\label{sec:model}

\subsection{Elliptical halo shear signals}
\label{sec:halo-model}

We use the results of \citet{acd15}, whose model for the effect of halo ellipticity on weak lensing observables will be useful in interpreting our measurements.
This model for the surface density of elliptical halos uses a multipole expansion:
\bea
\Sigma(R, \theta) & \propto & R^{\eta_0} [1 + \epsilon (-\eta_0/2) \cos{2\theta} + \mathcal{O}(\epsilon^2)] \\
 & \equiv & \Sigma_0 (R) + \Sigma_2 (R) \cos{2\theta} + ...
\eea
where $\eta_0 (R) = \de \log \Sigma_0 / \de \log R < 0$ and we assume the coeffecient of the quadrupole $-\epsilon \eta_0 / 2 \ll 1$, justifying the neglect of higher orders in the expansion.
Here $R$ is the projected distance from the center of the halo, and $\theta$ is the angle relative to the halo's major axis.
We set
\be \label{eq:assume}
\epsilon \approx \frac{-2 \Sigma_2 (R)}{\eta_0(R) \Sigma_0 (R)} \, ,
\ee
thus allowing the quadrupole $\Sigma_2$ to be completely determined by the monopole $\Sigma_0$, up to a proportionality factor $\epsilon$, the magnitude of the ellipticity.
This ellipticity is related to the axis ratio of the projected mass distribution, $q$, by $\epsilon = (1 - q^2) / (1+q^2)$.
Note that our ellipticity definition is bounded on the interval [0,1] and thus differs by a factor of 2 from \citet{acd15}.

\citet{acd15} shows that the tangential and cross-components of shear from the quadrupole are given by
\bea
\Sigma_{\rm crit} \, \gamma_+ & = & (\epsilon/2) \, [\Sigma_0 (R) \eta_0(R) - I_1 (R) - I_2(R)] \cos{2\theta} \label{eq:gplus} \\
\Sigma_{\rm crit} \, \gamma_\times & = & (\epsilon/2) \, [-I_1 (R) + I_2(R)] \sin{2\theta} \, , \label{eq:gcross}
\eea
where
\bea
I_1 (R) & \equiv & \frac{3}{R^4} \int_0^R R'^3 \Sigma_0 (R') \eta_0 (R') \de R' \, , \\
I_2 (R) & \equiv & \int_R^\infty \frac{\Sigma_0 (R') \eta_0 (R')}{R'} \de R' \, .
\eea
(The lensing weight function $\Sigma_{\rm crit}$ is defined later in Eq.~\ref{eq:sig-crit}.)
Note that throughout this paper we are concerned only with shear from the quadrupole: all shear variables $\gamma_+, \gamma_\times$, etc., refer to shear from the quadrupole and have no contribution from the monopole.

\begin{figure}
\centering
\resizebox{85mm}{!}{\includegraphics{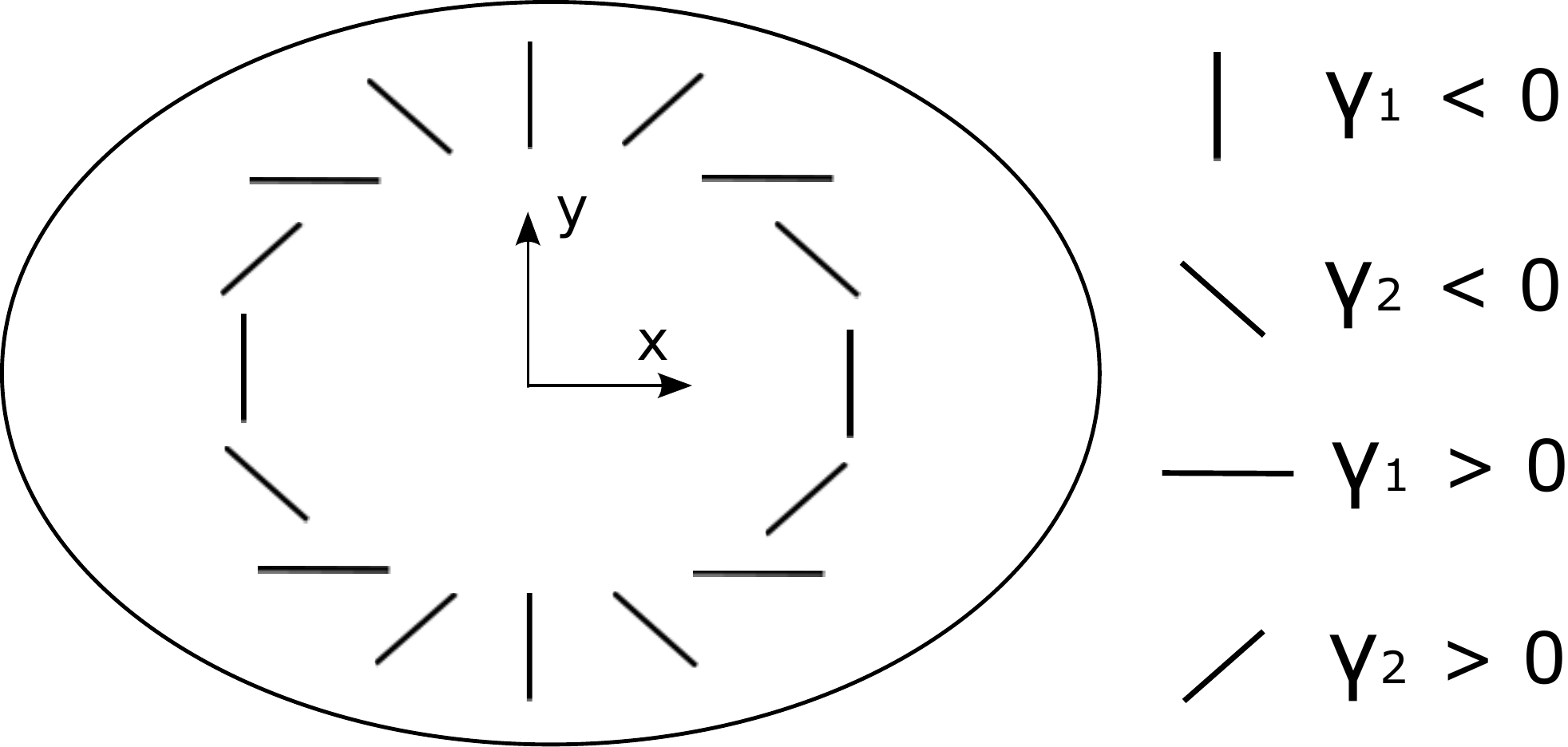}}
\caption{The quadrupole shear pattern produced by an elliptical dark matter halo. The x-axis of the Cartesian coordinate system is aligned with the major axis of the galaxy light distribution, assumed to be aligned with the major axis of the dark matter halo. Our sign convention for the Cartesian shear components is shown at the right.
Note that the monopole shear (which is purely tangential) is not contributing to the pictured shear pattern.
}
\label{fig:shear}
\end{figure}

In \S~\ref{sec:mandelbaum} we take a careful look at weighted estimators of tangential and cross shear that have been studied elsewhere in the literature \citep{nr2000,mhb06,vhs12}.
However, for most of this work we use Cartesian estimators, measured with respect to a coordinate system with the x-axis aligned with the lens LRG's major axis.
Our sign convention for these Cartesian $\gamma_1$ and $\gamma_2$ components is shown in Fig.~\ref{fig:shear}.
In this coordinate system, the two shear components are related to the tangential and cross-shear by
\bea
\gamma_1 (R,\theta) & = & -\gamma_+ (R,\theta) \cos{2\theta} + \gamma_\times (R,\theta) \sin{2\theta} \label{eq:gamma1a} \\
\Sigma_{\rm crit} \gamma_1 (R,\theta) & = & (\epsilon/4) \, [(2 I_1(R) -\Sigma_0(R) \eta_0(R)) \cos{4\theta} + \nonumber \\
 & & 2 I_2(R) - \Sigma_0 (R) \eta_0(R)] \, , \label{eq:gamma1}
\eea
and
\bea
\gamma_2 (R,\theta) & = & -\gamma_+ (R,\theta) \sin{2\theta} - \gamma_\times (R, \theta) \cos{2\theta} \label{eq:gamma2a} \\
\Sigma_{\rm crit} \gamma_2 (R,\theta) & = & (\epsilon/4) \, [2 I_1 (R) - \Sigma_0 (R) \eta_0(R)] \sin{4\theta} \, . \label{eq:gamma2}
\eea
Studying these equations, we see that the angular dependence of the shear goes as $\cos{4\theta}$ or $\sin{4\theta}$.
Thus there is a sign change in both components after every angle $\pi/4$, and moving around the circle the shear signal from elliptical halos transitions between regions where $\gamma_1$ then $\gamma_2$ alternately dominate.
The resulting shear pattern of Eqs.~(\ref{eq:gamma1}, \ref{eq:gamma2}) is sketched in Fig.~\ref{fig:shear}.

There are two minor exceptions to the shear direction of Fig.~\ref{fig:shear}: both are a result of the two terms without $\theta$ dependence in Eq.~(\ref{eq:gamma1}) for $\gamma_1$.
First, for a small part of our measurement range (below $R \sim 0.14 \mpch$, see Fig.~\ref{fig:cartesian}) the sign of $\gamma_1$ is negative for all $\theta$.
Second, the zero-crossing of $\gamma_1$ for different $\theta$ depends somewhat on scale $R$.
This leads to a small mixing of positive and negative shears in the separate ranges of $\theta$ we select below.
However, we have checked that this only leads to an appreciable loss of S/N in a small range around $R \sim 0.14 \mpch$: at scales $R < 0.1 \mpch$ and $R > 0.3 \mpch$ the loss of S/N due to mixing positive and negative shears in one bin is less than 10\%.
Both of these effects are taken into account in our modeling so that our constraints on halo ellipticity are not biased.
Note that while we are only concerned with the total signal from the halo quadrupole, the origin of the constant shear term has been studied by \citet{bn09} who determined that it comes from quadrupoles external to the source radius $R$.
(Note that at very small radii $R$, most of our elliptical halo is an ``external quadrupole,'' since this mass is outside the radius of interest.)
This distinction between terms coming from quadrupoles internal and external to the source radius may be useful in future higher signal-to-noise measurements, e.g., in discriminating between the quadrupole of the dark matter halo and that of correlated large-scale-structure.
Since our measurements focus on small scales, in this work we do not include a quadrupole from correlated large-scale-structure in the model.

These sign changes suggest grouping together the $\gamma_1$ component of background sources in the regions where $\cos(4\theta)$ has the same sign, i.e.,
\bea
\Delta\Sigma_1^{(-)} (R) & \equiv & \frac{4}{\pi} \int_{-\pi/8}^{\pi/8} \Sigma_{\rm crit} \gamma_1 (R,\theta) \, \de \theta \;\;\;\;\; {\rm and} \label{eq:dsom} \\
\Delta\Sigma_1^{(+)} (R) & \equiv & \frac{4}{\pi} \int_{\pi/8}^{3\pi/8} \Sigma_{\rm crit} \gamma_1 (R,\theta) \, \de \theta \, ,
\eea
where we have introduced the notation of $\Delta\Sigma$ in analogy to tangential shear, for which $\Delta\Sigma \equiv \Sigma_{\rm crit} \gamma_+$.
We do the same for the Cartesian cross-component, $\gamma_2$, except here we group together same-sign regions of $\sin(4\theta)$:
\bea
\Delta\Sigma_2^{(-)} (R) & \equiv & \frac{4}{\pi} \int_{0}^{\pi/4} \Sigma_{\rm crit} \gamma_2 (R,\theta) \, \de \theta \, , \\
\Delta\Sigma_2^{(+)} (R) & \equiv & \frac{4}{\pi} \int_{\pi/4}^{\pi/2} \Sigma_{\rm crit} \gamma_2 (R,\theta) \, \de \theta \, . \label{eq:dstp}
\eea
In order to capture the information on halo ellipticity in the shear field, we measure these four functions from the data.
Note that the above integrals only cover 1/4 of the circle for each shear component, but including the other regions offset by an angle $\pi/2$ gives equivalent results, as can be seen in Fig.~\ref{fig:shear}.
When measuring these functions, in order to increase statistics we exploit these symmetries and average together all regions with the same sign of the shear (see the estimators in \S~\ref{sec:measure}).

These estimators also have the useful property that purely tangential shears (the lensing monopole) are nulled automatically.
This can be seen from Fig.~\ref{fig:shear}:
consider two source galaxies located at $(x,y)$ positions $(R,0)$ and $(0,R)$, i.e., in the regions where we measure $\dsom$ as in Eq.~(\ref{eq:dsom}).
Both have monopole contributions of $\gamma_t > 0$ which are purely tangential and equal in magnitude (since they are located at the same distance $R$ from the LRG center).
Transforming to Cartesian shears $(\gamma_1, \gamma_2)$, the two sources have $(-\gamma_t, 0)$ and $(\gamma_t, 0)$, respectively.
When averaged in the same bin, the net monopole contribution from these two sources is zero.
More generally, whenever two sources are offset by 90$^{\circ}$, relative to the LRG center, the Cartesian components of the monopole shear cancel.
Assuming background sources are distributed isotropically around the lens, each source used in any of the four estimators in Eqs.~(\ref{eq:dsom}-\ref{eq:dstp}) will always have a paired source offset by 90$^\circ$.
(Note that we have checked that our sources are indeed distributed isotropically.)
For more detailed discussion of this nulling of monopole shear, and its usefulness in constructing an estimator of filament lensing, see \citet{clampitt16}.

Note that in the absence of systematics, the form of Eqs.~(\ref{eq:gamma1}, \ref{eq:gamma2}) suggests estimators weighted by $\sin{4\theta}$ and $\cos{4\theta}$. The weighting is not  optimal since it actually loses signal in $\gamma_1$ by averaging out the terms in Eq.~(\ref{eq:gamma1}) that have no angular dependence.
In addition,  the presence of significant galaxy-halo misalignment (\S~\ref{sec:align}) introduces an error in the choice of the angle $\theta$ for each lens-source pair and dilutes the stacked signal. We find that for our sample the simple $\sin{4\theta}$ and $\cos{4\theta}$ weighting does not lead to an improvement in the signal-to-noise (see \S~\ref{sec:misc}). We  leave for future work the exploration of estimators that may improve on the ones presented here. 

Our model predicts the shear quadrupole from the monopole, up to an overall factor of the ellipticity.
Thus we use the parameters of these LRG halos which have been determined by \citet{msh08} using isotropic galaxy-galaxy lensing: the average mass and concentration are $M = 4\times10^{13} M_\odot/h$ and $c=5$.
These parameters, along with our assumed NFW halo profile \citep{nfw1997} then fully determine the signals of monopole $\Sigma_0 (R)$ and quadrupole, Eqs.~(\ref{eq:dsom})-(\ref{eq:dstp}). We discuss below the impact of varying $M$ and $c$ on our ellipticity estimate.

 \subsection{Systematic shear}
 \label{sec:syst}

For the SDSS survey, \citet{mhb06} found that nearby galaxy shapes tend to be slightly aligned with each other.
There are several factors contributing to this alignment.
First, the PSF has a net large-scale alignment with the survey scan direction, and any incomplete correction of this PSF will tend to affect nearby galaxies in the same way.
In addition, \citet{hhm14} found that specific declination ranges have systematic shear due to an incorrect PSF model in the r-band camcol 2 CCD; see especially the discussion around their Fig.~8.
(Note also Sec.~4.4.3 of \citet{rmg12}, which studied the effect of this systematic on the shear catalog of \citet{mhb06} in more detail.)
Both effects imprint a slight alignment between our lens and source ellipticities.
For isotropic galaxy-galaxy lensing, these systematics are not a worry because they are largely cancelled out along with other additive errors.

One significant advantage of using the Cartesian shear components described above to measure halo ellipticity is that this major systematic error only affects the $\Delta\Sigma_1^{(+/-)}$ functions, leaving $\Delta\Sigma_2^{(+/-)}$ untouched.
Once one averages over many lens-source pairs, the only preferred orientation is the major/minor axis of the lens which  are aligned with the x- and y-axes. A systematic error can contribute to the estimated shear along these axes, and thus contaminate the 1-component, as discussed below for SDSS. However by symmetry, the net contribution of a systematic to the 2-component of the shear must vanish since both signs are equally likely, unlike the quadrupole signal. 
Previous attempts at measuring halo ellipticity used the tangential and cross-components of the shear, which can 
both be contaminted by  systematic errors. One way to see this is the 1-component has non-zero projections along both the tangential and cross-components (except at special points such as $\theta=0$).
Note that in addition to these arguments from symmetry, in Fig.~\ref{fig:foreground} we have tested independently that $\Delta\Sigma_2^{(+/-)}$ has no contribution from this systematic.

The elliptical halo shear measurement of \citet{mhb06} is dominated by this systematic for at least one lens sample (the red ``L4'' lenses in Fig.~7 of that work).
For this sample, the systematic has a roughly $1/R$ power law dependence.
Thus we parameterize the systematic shear as
\be \label{eq:syst}
\Delta\Sigma^{\rm syst} \propto \frac{1}{R}
\ee
and marginalize over the amplitude, which is required to be positive according to our shear convention.
This can be seen from Fig.~\ref{fig:shear}, where an alignment of the lens galaxy and source whiskers will contribute positively to $\gamma_1$.
We have also tried other power law models, such as $1/R^2$, for the systematic shear and find our results are not sensitive to the power chosen.
Our full model for the $\Delta\Sigma_1$ components is therefore $\Delta\Sigma_1^{(+/-)} + \Delta\Sigma^{\rm syst}$.
Note that while our main results involve modeling and marginalizing over the systematic, in \S~\ref{sec:syst2} and Fig.~\ref{fig:foreground} we compare with model-independent tests of the systematic shear and find similar results.

Finally, we comment briefly on the possibility of a systematic anti-alignment of lenses and sources (negative amplitude for Eq.~(\ref{eq:syst})).
The incorrect PSF model in the r-band camcol 2 CCD is a large effect that always acts to align nearby galaxies.
For the other camcols, since lenses and sources have different magnitudes and sizes, one could imagine the possibility that either the lens or source would be overcorrected and the other undercorrected.
Thus, when performing the fits we will also test the results when dropping the requirement that the systematic amplitude be positive.

\section{Data and Measurement Method}
\label{sec:measure}

\begin{figure*}
\centering
\resizebox{180mm}{!}{\includegraphics{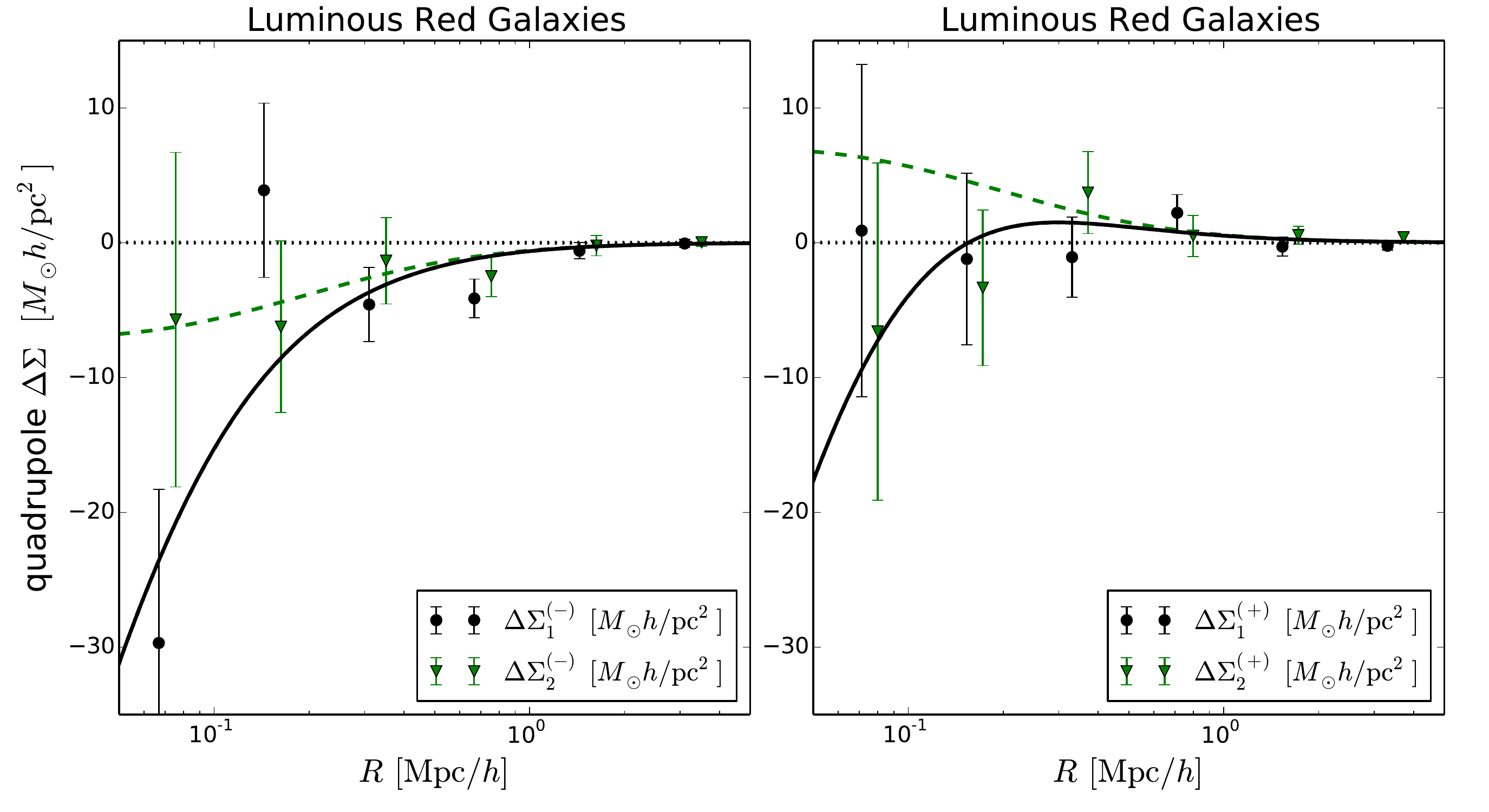}}
\caption{({\it Left panel}): The measured values of $\dsom$ (black points) and best-fit elliptical halo model (solid line), along with the same for $\dstm$ (green triangles and dashed line).
({\it Right panel}): Same as the left panel, but showing the data and predictions for $\Delta\Sigma^{(+)}_{1,2}$.
For all cases, the model is able to describe the data very well.
As predicted from Eqs.~(\ref{eq:gamma1}) and (\ref{eq:gamma2}), the constant (no angular variation) shear term causes $\dsom$ to have the largest signal.
The best fit value of ellipticity $\epsilon$ is 0.24, while the systematic shear in our model has a best-fit value which is close to zero for all scales $R$ on this plot.
}
\label{fig:cartesian}
\end{figure*}

We use shear catalogs and photometric redshift probability distributions that are almost identical to those in \citet{sjs2009} and \citet{scm12}, respectively.
The catalog contains 34.5 million sources, with a redshift distribution peaking at $z \sim 0.35$ and a long tail towards higher redshifts.
In order to find the major axes of the LRG light distributions, we match the shear catalog of \citet{sjs2009} to the DR7-Full LRG catalog of \citet{kbs10}.
This is necessary to align our Cartesian coordinate systems for each lens prior to performing a stacked lensing measurement.
This results in 69,286 matching galaxies with redshifts $0.16 < z < 0.47$ in our lens sample.
Essentially all of these matched LRGs have quality shear measurements, based on the small shape measurement error of each galaxy.
See \citet{kbs10} for more details on the LRG catalog, \citet{sjs2009} for more details on the shear catalog, and \citet{sjf2004} for more detailed descriptions of the shear measurement method.
Note that although the discussion of systematic shear in Sec.~\ref{sec:syst} focuses on the shear catalog of \citet{rmg12}, the same PSF models were used for our shear catalog (Erin Sheldon and Rachel Mandelbaum, private communication).
Since the source of systematics was primarily the PSF models \citep{hhm14,rmg12}, the results are also relevant for our work.

To perform the measurement, we begin by rotating the \texttt{RA} and \texttt{DEC} coordinates of all nearby background galaxies into a Cartesian coordinate system with $x$-axis along the LRG major axis (as pictured in Fig.~\ref{fig:shear}).
The source catalog has shear components $(\gamma'_1, \gamma'_2)$ which are also defined with respect to the \texttt{RA}, \texttt{DEC} coordinate system.
We simultaneously rotate these into ($\gamma_1, \gamma_2$) components defined with respect to the $(x,y)$ system, keeping in mind that shears rotate twice as fast as galaxy positions ($\gamma = \gamma' e^{2i\theta'}$ for a rotation by $\theta'$).
Our lensing observable is then given by
\begin{equation}
\widehat{\Delta\Sigma}_{k}^{(s)} (R)=\frac{\sum_j \left[ w_j \left(\ave{\Sigma_{\rm crit}^{-1}}_j(\zl)\right)^{-1}
 \hat{\gamma}_{k,j}(R) \right]}{\sum_j w_j}
\label{eq:dsig}
\end{equation}
where the sum goes over all sources $j$ that fall in the specified region, different for each shear component $k$ and sign $s$:
\bea
k = 1, \, s = + ; & 0 \le (\theta_j + \pi/8) < \pi/4 \nonumber \\
k = 1, \, s = - ;  &\pi/4 \le (\theta_j + \pi/8) < \pi/2 \nonumber \\
k = 2, \, s = + ; & 0 \le \theta_j < \pi/4 \nonumber \\
k = 2, \, s = - ; & \pi/4 \le \theta_j < \pi/2
\eea
and where each estimator also includes the sources in symmetrical regions shifted by $\pi/2$, $\pi$, and $3\pi/2$, as shown in Fig.~\ref{fig:shear}.
Note that the Cartesian shear of the j'th source galaxy, $\hat{\gamma}_{k,j}$, should be distinguished from the predicted shears $\gamma_1$ and $\gamma_2$ referred to in our model Eqs.~(\ref{eq:gplus}-\ref{eq:dstp}).
The summation $\sum_j$ runs over all the background galaxies
in the radial bin $R$, around all the LRG positions.
The weight for the $j$-th galaxy is given by
\be \label{eq:weight}
w_j = \frac{\left[
\ave{\Sigma_{{\rm crit}}^{-1}}_j(\zl)\right]^{2}}
{\sigma_{\rm shape}^2 + \sigma_{{\rm meas},j}^2}.
\ee
The method for these weights follows that used for tangential shear measurements \citep{msh08b, scm12, msb2013}.
We use $\sigma_{\rm shape} = 0.32$ for the rms intrinsic ellipticity  and $\sigma_{{\rm meas},j}$ denotes measurement noise on each background galaxy ellipticity. 
$\ave{\Sigma_{\rm crit}^{-1}}_j$ is the lensing critical density for the $j$-th source galaxy, computed by taking into account the photometric redshift uncertainty:
\be
\ave{\Sigma_{{\rm crit}}^{-1}}_j(\zl)
= \int_0^\infty\! {\rm d}\zs \Sigma_{\rm crit}^{-1} (\zl, \zs) P_j(\zs),
\ee
where $\zl$ is the redshift of the LRG and $P_j(\zs)$ is the probability distribution of the photometric redshift for the $j$-th galaxy.
Note that $\Sigma_{\rm crit}^{-1}(\zl,\zs)$ is computed as a function of lens and source redshifts for the assumed cosmology as
\be \label{eq:sig-crit}
\Sigma_{\rm crit} (\zl, \zs) = \frac{c^2}{4\pi G} \frac{D_A(\zs)}{D_A(\zl) D_A(\zl,\zs)} \, ,
\ee
and we set $\Sigma_{\rm crit}^{-1}(\zl,\zs)=0$ for $\zs<\zl$ in the computation.
The angular diameter distances from the observer to source, observer to lens, and lens to source are given by $D_A(\zs)$, $D_A(\zl)$, and $D_A(\zl, \zs)$, respectively.

The error bars and covariance are obtained with the jackknife method of \citet{nbg09}, using a set of 134 jackknife patches.
Note that given the size of the error bars in the following section, a few percent uncertainty in the multiplicative bias is not significant.
In addition, the effects of additive bias will be treated in detail in \S~\ref{sec:syst2}.
Thus we directly use the jackknife covariances in all plots and fits.

\section{Results}

\subsection{Measurement with Cartesian components}

The results of the measurement and best-fit model prediction are shown in Fig.~\ref{fig:cartesian}.
At $R \sim 0.3 \mpch$ the magnitude of the measured signal is $\sim 5 \mhpc$ for $\dsom$.
Comparing to measurements of the monopole signal for this LRG sample \citep{msh08, msb2013}, the magnitude of each component of the quadrupole is $\sim 10$ times smaller.
By $2\mpch$ the measured signal is consistent with zero.
Since this is  the transition scale between the 1- and 2-halo regimes \citep{vvh14} for halos of mass $\sim 4\times10^{13} M_\odot/h $, our simple 1-halo modeling of $\Sigma_0 (R)$ is sufficient to describe the signal for the purposes of this study.
However, more detailed models will be needed with higher signal-to-noise measurements from upcoming survey data.

The other components are predicted to have lesser overall signal-to-noise (S/N) than $\dsom$, and this is consistent with what we see in the data.
Since the noise in all components are expected to be dominated by shape noise at these scales, and each has the same shape noise, the components with lesser predicted signal should also have lesser S/N.
The predicted $\dstm$ and $\dstp$ have identical amplitude and lower S/N, and $\dsop$ has the least.

\begin{figure}
\centering
\resizebox{78mm}{!}{\includegraphics{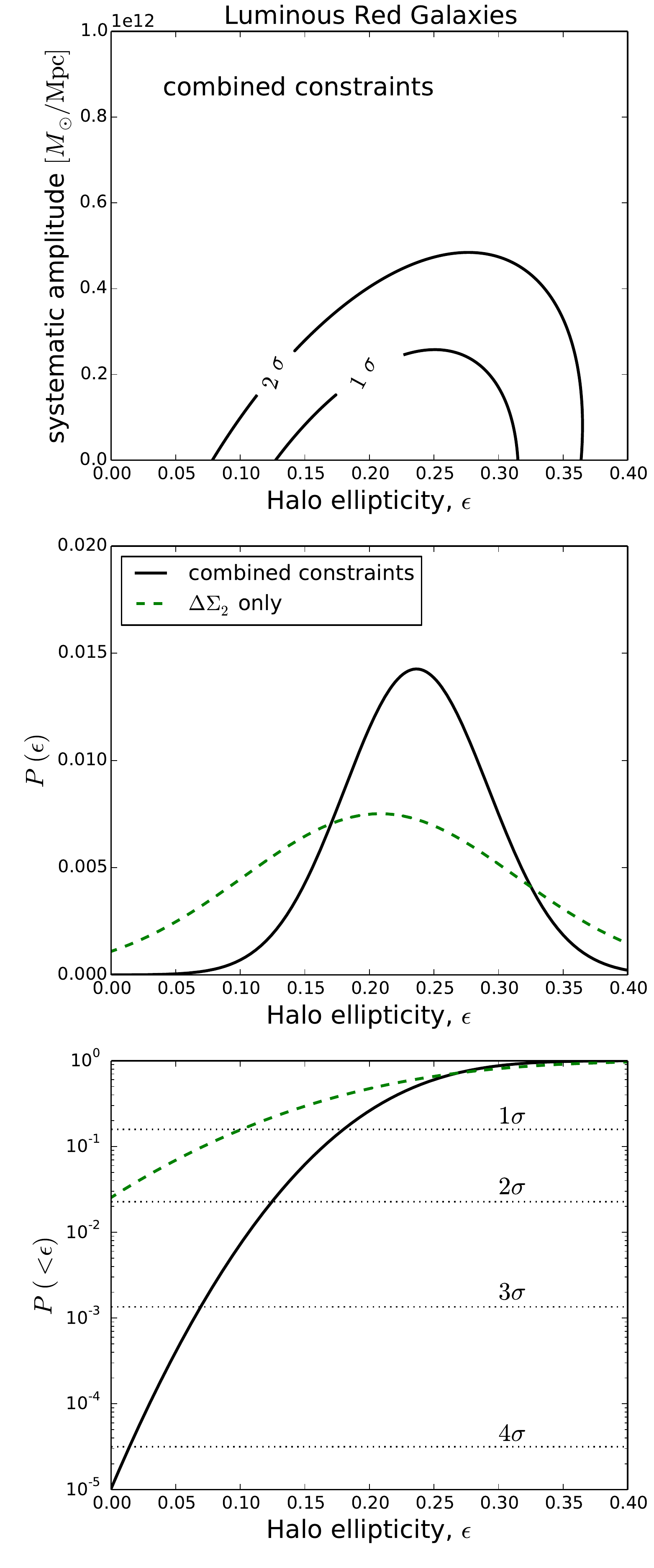}}
\caption{({\it Top panel}): Combined constraints from all four functions $\Delta\Sigma^{(+/-)}_{1,2}$ on the halo ellipticity $\epsilon$ and systematic shear amplitude. For allowed values within the $2\sigma$ contours, the magnitude of the systematic is at most 1\% of the signal over all measured scales.
({\it Middle panel}): Probability of the halo ellipticity given the data, marginalized over the systematic amplitude.
Constraints are pictured for our entire data vector (solid line) and only the components that are not dependent on the systematic model, $\Delta\Sigma^{(+/-)}_2$ (dashed line).
({\it Lower panel}): Cumulative probability distribution of the middle panel, corresponding to the p-value of a given ellipticity $\epsilon$.
The $1$-$4\sigma$ confidence levels (horizontal dotted lines) show that we rule out the hypothesis of no halo ellipticity $\epsilon$ at $4\sigma$, and $\epsilon < 0.12$ at $2\sigma$.
These lower bounds hold even in the presence of misalignment between the light and dark matter profiles.
}
\label{fig:epsilon}
\end{figure}

\subsection{Constraints on halo ellipticity}
\label{sec:constraints}

In Fig.~\ref{fig:epsilon} we show our constraints on the halo ellipticity and the systematic alignment amplitude.
The 2D constraints in the top panel show the systematic amplitude (see Eq.~\ref{eq:syst}) is very small for our LRG lens sample, with a best-fit amplitude less than $\sim 0.1 \times 10^{12} M_\odot/{\rm Mpc}$.
Thus, at $R = 1 \mpch$, the systematic has magnitude less than $0.1 \, \mhpc$ in units of $\Delta\Sigma$, and at all scales it is much smaller than our modeled elliptical halo signal.
Marginalizing over the amplitude of the systematic, we rule out the hypothesis of no ellipticity $\epsilon$ at $4\sigma$, and $\epsilon < 0.12$ at $2\sigma$.
These constraints can be read off from the lower panel, which shows the cumulative probability in the low $\epsilon$ tail of the posterior.
Possible misalignment between the galaxy and dark matter distributions only dilutes our measurement (see the discussion in \S~\ref{sec:align}), so these lower bounds on halo ellipticity still hold in the presence of misalignment.
The best-fit model has $\epsilon = 0.24\pm0.06$.
(Note that 0.24 refers to the mean of the posterior and $0.06$ is the $1\sigma$ confidence interval.)
With a reduced $\chi^2$ of 25.1/22, the $\epsilon=0.24$ model is a good fit to the data.
In contrast, taking the null hypothesis model of no halo ellipticity the reduced $\chi^2$ is 41.7/24, a very poor fit.
It should be noted that this best-fit value may be an underestimate due to misalignment of the light and dark matter position angles.
We discuss this possibility in detail in \S~\ref{sec:align}.
Note that the covariance between the 24 data points in Fig.~\ref{fig:cartesian} is small but non-negligible: thus the full covariance has been used in these constraints.

Also shown in Fig.~\ref{fig:epsilon} is the constraint from the $\Delta\Sigma_2^{(+/-)}$ observable alone, i.e., not including the higher signal-to-noise $\Delta\Sigma_1^{(+/-)}$ constraints.
Even if our model for the systematic is somehow flawed, this observable is unaffected by the systematic and it gives a $2\sigma$ constraint by itself.
In \S~\ref{sec:syst} we discussed the motivation for not allowing a systematic anti-alignment (which will be somewhat degenerate with the constant-with-angle terms of Eq.~(\ref{eq:gamma1})) in our fits.
Nonetheless, if we redo the fits allowing a negative amplitude of Eq.~(\ref{eq:syst}), the mean of the systematic amplitude is then $\sim (-0.35 \pm 0.4) \times 10^{12} M_\odot/{\rm Mpc}$, still consistent with zero.
Furthermore, given our tests in \S~\ref{sec:syst2} with a foreground ``source'' sample that has no lensing contribution, we think it unlikely that the systematic produces a net anti-alignment.
In the following section, we repeat the measurement using estimators previously used in the literature, and make comparison to previous results of \citet{mhb06}.

\begin{figure}
\centering
\resizebox{85mm}{!}{\includegraphics{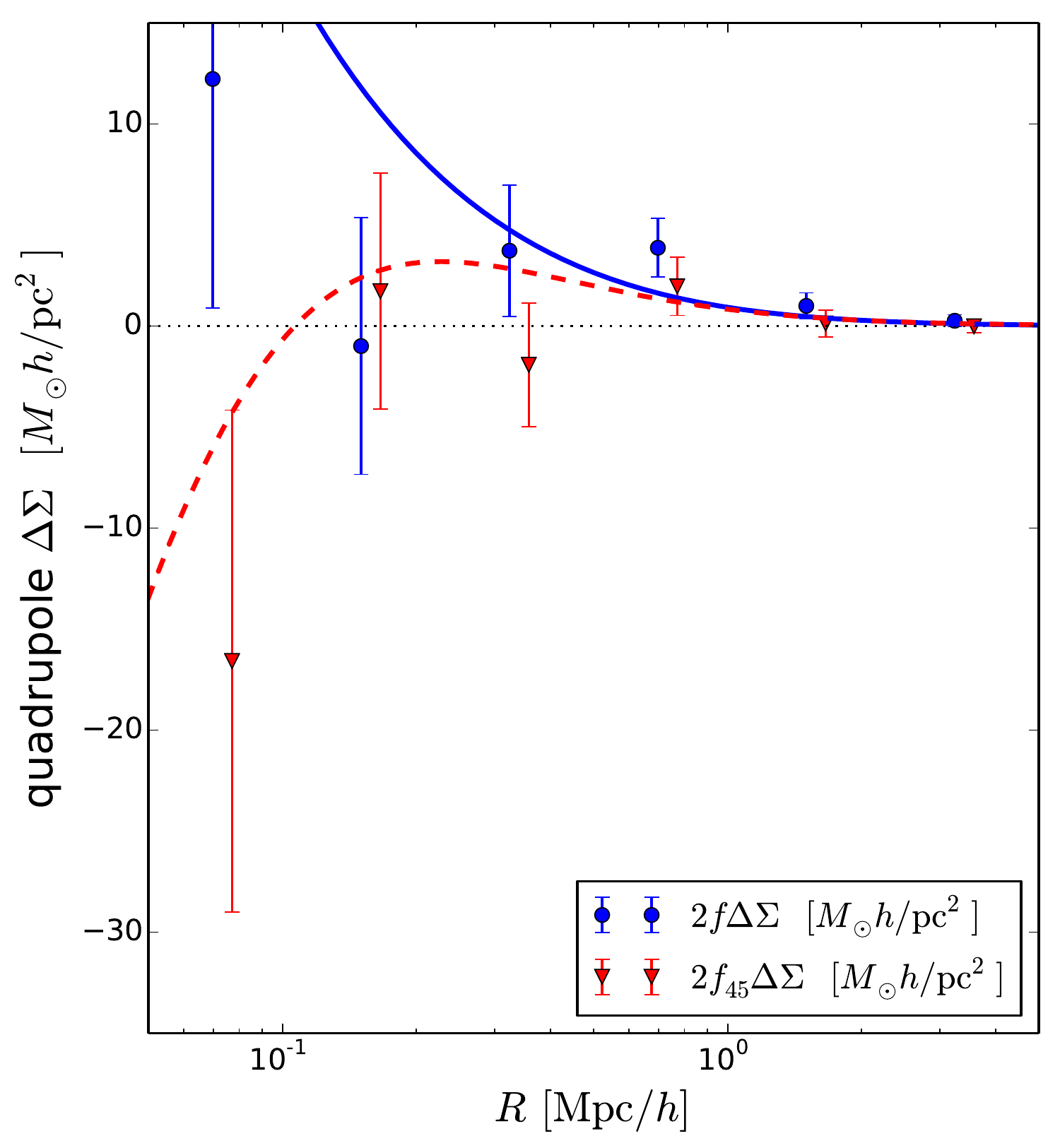}}
\caption{The tangential- (blue points) and cross-shear (red triangles) estimators used in \citet{mhb06}, as well as the corresonding best-fit model prediction to each (solid and dashed lines, respectively).
These also give a significant signal with our lens and source sample, but subtraction of the two components to deal with systematic shear will also remove real signal from the elliptical halo.
}
\label{fig:tan}
\end{figure}

\subsection{Measurement with tangential components}
\label{sec:mandelbaum}

\begin{figure*}
\centering
\resizebox{180mm}{!}{\includegraphics{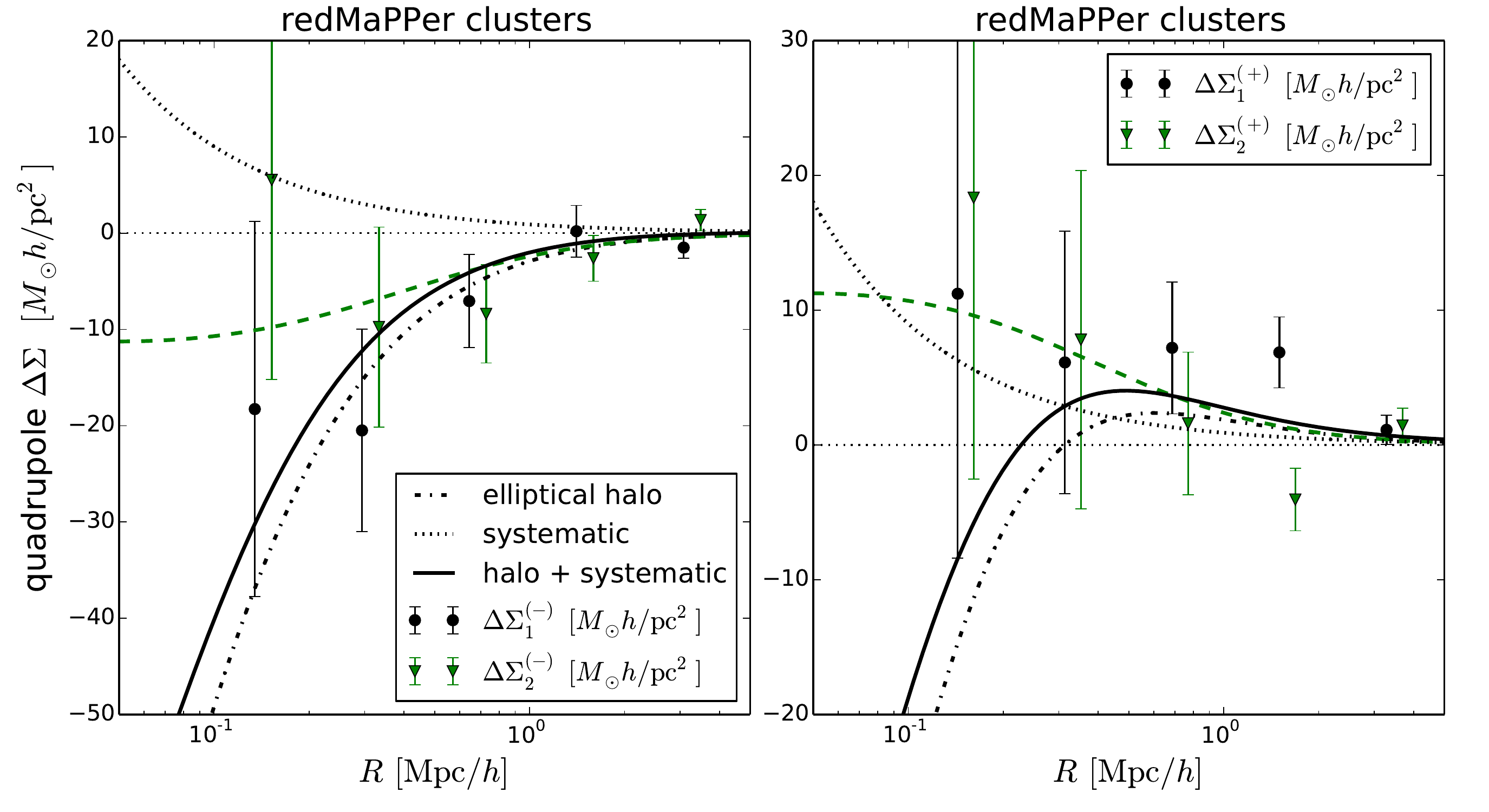}}
\caption{Same as Fig.~\ref{fig:cartesian}, but for redMaPPer clusters.
The measurement is still signal dominated, but for these clusters the systematic (dotted line) is not negligible with an amplitude $\sim 20-40\%$ of the halo ellipticity signal (dash-dot line).
The total signal (solid and dashed lines) is 2-3 times larger than for the LRG sample.
}
\label{fig:redmap}
\end{figure*}

The following estimators were developed by \citet{nr2000} and applied to survey data by \citet{mhb06}:
\be \label{eq:cos}
f \Delta \Sigma (R) = \frac{\sum_j w_j \left(\ave{\Sigma_{\rm crit}^{-1}}_j(\zl)\right)^{-1} \hat{\gamma}_{+,j} \cos{2\theta_j}}{2\sum_j w_j \cos^2{2\theta_j}}
\ee
for the tangential component of shear and
\be \label{eq:sin}
f_{45} \Delta \Sigma (R) = \frac{\sum_j w_j  \left(\ave{\Sigma_{\rm crit}^{-1}}_j(\zl)\right)^{-1} \hat{\gamma}_{\times,j} \sin{2\theta_j}}{2\sum_j w_j \sin^2{2\theta_j}}
\ee
for the cross-component.
In this section we repeat the measurements on our lens and source samples using these estimators, and compare the results with \citet{mhb06}.
Given our signal in Fig.~\ref{fig:cartesian}, we would expect both estimators to give positive signals when repeated using our lens and source samples.
Indeed, this is what we see in Fig.~\ref{fig:tan} for $f \Delta\Sigma$, while the results for $f_{45} \Delta\Sigma$ are more noisy.
The positive sign is partly a matter of sign conventions: ours are given by Fig.~\ref{fig:shear} and Eqs.~(\ref{eq:gamma1a}, \ref{eq:gamma2a}).
Note that our Eqs.~(\ref{eq:cos}) and (\ref{eq:sin}) include an extra factor of 2 in the denominator relative to Eq.~(\ref{eq:dsig}) in order to match the convention of \citet{mhb06}.
In Fig.~\ref{fig:tan} we plot $2 f\Delta\Sigma$ and $2f_{45} \Delta\Sigma$ so that the signal amplitude can be compared fairly with our fiducial measurements in Fig.~\ref{fig:cartesian}.
\citet{mhb06} also included a factor of the lens ellipticity in the estimators Eqs.~(\ref{eq:cos}, \ref{eq:sin}); we have found that including this factor has a negligible effect on the measurement.
We have chosen to leave it out for simplicity and to better match the form of our own estimator equations.

The most comparable lens sample of \citet{mhb06} is the red ``L6'' lenses, for which they find an $\sim 2-3\sigma$ hint of alignment between the elliptical halo and lens galaxy light distribution.
Our stronger detection is likely due to a few effects.
While their sample has $\sim$ 50\% more galaxies than our LRG sample ($\sim$ 107,000 compared to $\sim$ 70,000), the lack of spectroscopic redshifts likely dilutes their signal somewhat.
Perhaps more importantly, our estimator does not sacrifice signal in order to remove systematics.
\citet{mhb06} argue that both estimators have an equal contribution from the systematic shear, and so they use the difference $f \Delta\Sigma - f_{45} \Delta \Sigma$ to produce a halo ellipticity measurement that is free of this systematic.
They also note that this subtraction is not ideal, however, as there is also signal in $f_{45} \Delta \Sigma$ (see Fig.~2 of that work).
When subtracting the two estimators, the systematic will be cancelled, but some real signal will also be removed at the same time.
This can be seen in Fig.~\ref{fig:tan}, where above $R \sim 0.1 \mpch$ both estimators have positive signal from the elliptical halo.
At scales $\sim 0.5 \mpch$ and larger, the two have similar magnitudes as well, so that subtraction will remove most of the real signal as well as the systematic shear.
Thus we continue to focus on the Cartesian shear estimators of halo ellipticity in this paper.

\subsection{Measurement with redMaPPer Clusters}
\label{sec:redmapper}

As a check of our measurement method and interpretation, we attempt the same measurement with redMaPPer clusters \citep{rrb14, rrb15}.
We consider only those clusters where the most-probable central galaxy has a high probability of being the true center $(P_{\rm cen} > 90\%)$.
In addition, we use only clusters with redshift $0.1 < z < 0.3$, removing high-redshift clusters with noisier detections.
We then match these galaxies to our shear catalog, yielding the direction of the galaxy major axis.
Of the original $\sim$ 26,000 redMaPPer clusters, only $\sim$ 2,700 remain after these three cuts.

As for the LRG sample, we assume the cluster density follows an NFW profile and the quadrupole of the surface density is completely determined from the monopole via Eq.~(\ref{eq:assume}), up to the ellipticity amplitude.
We obtain an approximate mass distribution for this cluster sample via the mass-richness relation of \citet{rkr12},
\be
M_{\rm 200} = 10^{14} M_\odot/h \times \exp{(1.72 + 1.08 \ln{(\lambda / 60.)})} \, ,
\ee
where $\lambda$ is the cluster richness.
We expect this relation to only be a rough approximation for our sample: it was derived for maxBCG clusters and involved assuming a cosmology in order to apply abundance matching techniques.
The mean mass is $M_{200} \approx 3\times 10^{14} M_\odot/h$, roughly an order of magnitude larger than the mean mass of our LRGs.
To obtain the concentration we use the mass-concentration relation of \citet{dsk08}, yielding $c \approx 4.7$ at the mean mass.
Finally, the mean redshift of the sample is $z \approx 0.23$.
Note that many of these redMaPPer centrals are in fact LRGs \citep{hll15}: this measurement is therefore not completely independent from our LRG measurements.
LRGs occupy a wide range of halo masses, and those which are also redMaPPer centrals will occupy some of the largest halos.
Since the lensing quadrupole is expected to be proportional to the monopole, this measurement of cluster quadrupoles should have a higher signal than the LRG measurement: as will be shown below, the data matches this prediction.

The result of the measurement and best-fit model is shown in Fig.~\ref{fig:redmap}.
The constraints on the systematic shear permit a larger systematic amplitude than for LRGs, but the measurement is still signal-dominated.
For example, at $R\sim 0.1 \mpch$ the systematic amplitude is 20\% (35\%) of the halo ellipticity signal for the $\dsom$ ($\dsop$) estimator.
Reassuringly, the amplitudes of $\dstm$ and $\dstp$ are similar to each other, as expected if this shear component is insensitive to this systematic.

The expected amplitude of the lensing signal is larger for this sample than the LRGs, due to the larger mean mass of clusters.
In the absence of systematics such as misalignment of light and dark matter and correlation of lens and source ellipticities, we would expect the measured signal for clusters to be $\sim 2-3$ times larger, due to the 10 times larger mean mass.
(Recall that Eq.~(\ref{eq:assume}) requires the quadrupole to be proportional to the monopole, which we model with an NFW profile.)
The observed difference is close to this, but the detailed fits of ellipticity show a slightly smaller value for the clusters.
Although higher signal-to-noise measurements will be needed to tell whether this ellipticity difference is significant, it is plausible that these systematics are stronger for clusters.
With our cut $P_{\rm cen} > 90\%$, the redMaPPer galaxies should mostly be well-centered within the halo; nonetheless, it is still hard to compete with the LRG sample for which there is almost always only one very bright galaxy, making the halo center very easy to determine.
Also, it may be that the cluster halo's major axis correlates better with the major axis of the member galaxy distribution.
However, this is a very different method susceptible to different systematics \citep{eb09}, and so we leave it for future work.

\begin{figure}
\centering
\resizebox{85mm}{!}{\includegraphics{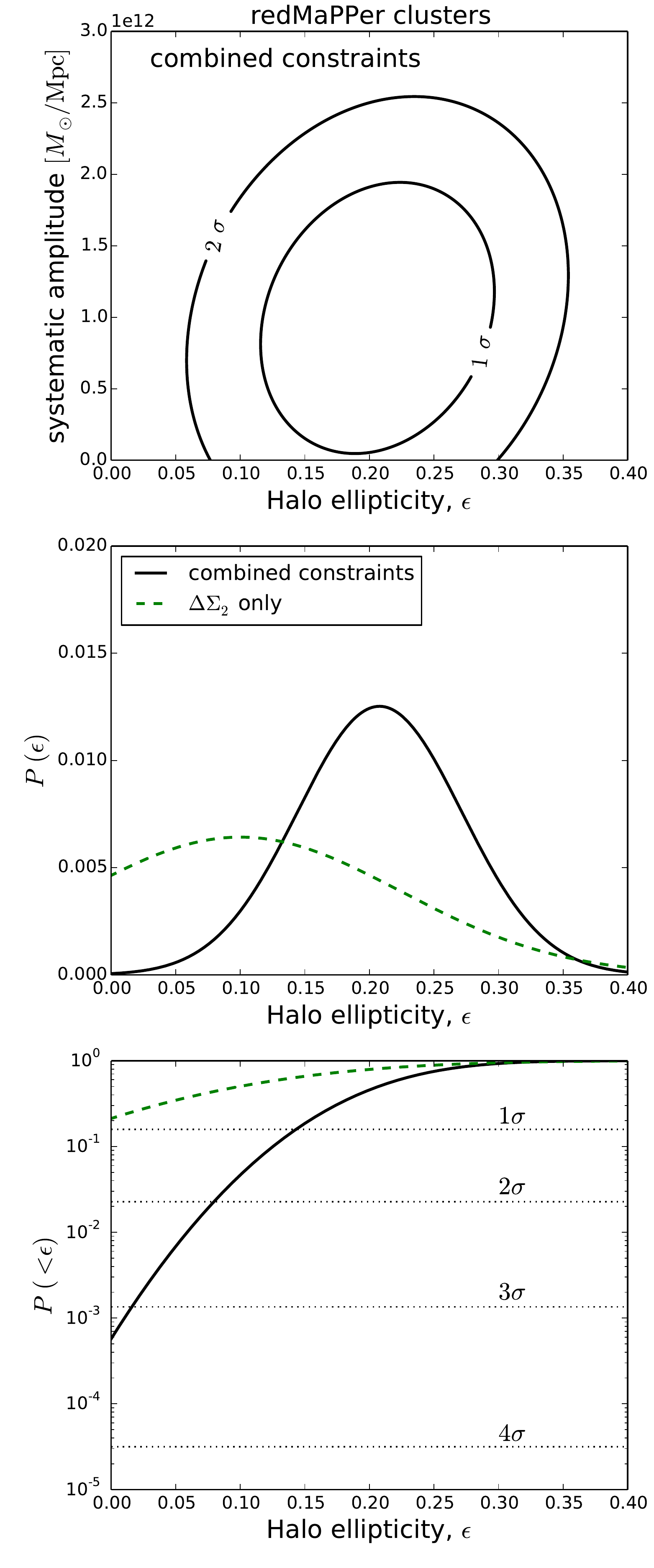}}
\caption{Same as Fig.~\ref{fig:epsilon}, but for redMaPPer clusters.
With a factor of 20 fewer objects, the noise of this cluster measurement is considerably larger, but the significance is still above $3\sigma$ due to the larger signal.
}
\label{fig:fit-redmap}
\end{figure}

Our fits of the systematic shear (top panel of Fig.~\ref{fig:fit-redmap}), show that this systematic is larger ($1-2\sigma$ significance) for redMaPPer clusters than LRGs.
The fits for halo ellipticity $\epsilon$ are also shown in Fig.~\ref{fig:fit-redmap}.
The best-fit ellipticity is $\epsilon = 0.21$, corresponding to an axis ratio 0.81.
The reduced $\chi^2$ of the best-fit model is $20.6/18$, an acceptable fit to the data.
The hypothesis of no halo ellipticity is ruled out at slightly larger than the $3\sigma$ confidence level.
The main reason for the smaller detection significance is that the number of clusters is less than 5\% of the number of LRGs.
As discussed above, the best-fit ellipticity may be smaller than LRGs due to greater misalignment of light and dark matter, but we leave a detailed study of major axis estimators for future work.

\section{Systematic and Other Checks}

\subsection{Systematic Lens-Source Alignment}
\label{sec:syst2}

Since the systematic described in \S~\ref{sec:syst} is due to camera and PSF related effects, it should be present regardless of whether a given source galaxy is at higher redshift than the LRG lens.
This suggests a model independent test of the systematic alignment between lenses and sources, in which we only use foreground galaxies as ``sources.''
Such galaxies will not be lensed and yet should be affected similarly by the systematic, providing a clean measurement of the systematic alignment.
We perform the measurement using all foreground sources with redshifts $\langle z_{\rm s} \rangle < z_{\rm L} - 0.1$, where $\langle z_{\rm s} \rangle$ is the mean of the individual source's redshift distribution $P(z_{\rm s})$.
Note that since the $\Sigma_{\rm crit}$ weighting is irrelevant for sources with redshift below the lens redshift, we drop this weighting and measure $\gamma$ instead of $\Delta\Sigma$.
One qualification in interpreting this test is that changing the redshift cut on the ``source'' sample can also change other mean properties such as size and surface brightness, which correlate with the level of PSF related systematics.
Furthermore, high redshift LRGs will be weighted more highly in this test than in the actual measurement, since they have more low redshift ``sources'' available.

\begin{figure}
\centering
\resizebox{85mm}{!}{\includegraphics{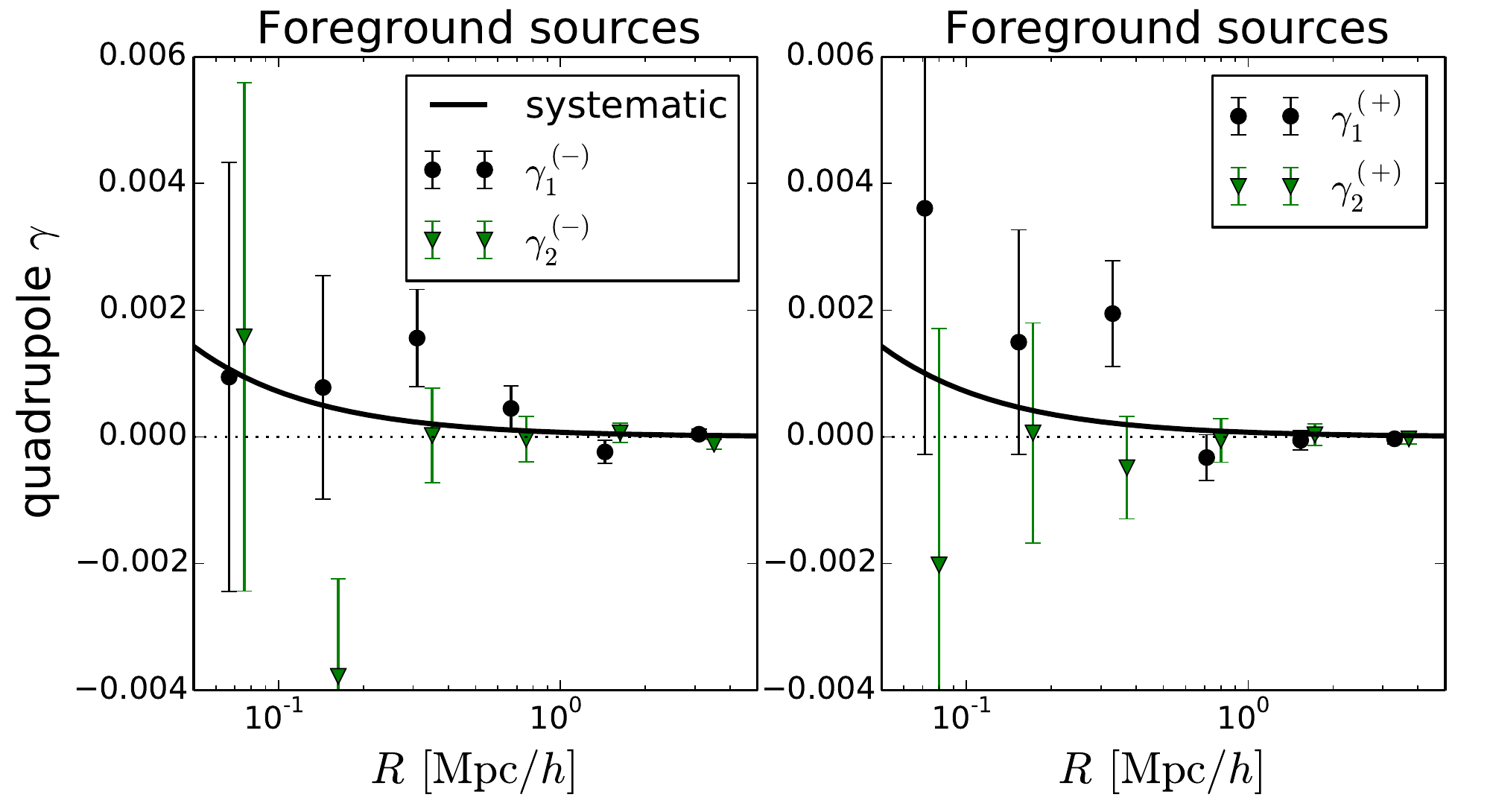}}
\caption{Same as Fig.~\ref{fig:cartesian}, but uses foreground sources which are not lensed to measure systematic alignment of LRGs and foreground galaxies.
The best-fit power law systematic (solid line) is fit to the $\Delta\Sigma_1$ points, while $\Delta\Sigma_2$ is consistent with zero.
}
\label{fig:foreground}
\end{figure}

The results are shown in Fig.~\ref{fig:foreground}.
Qualitatively it is consistent with our chosen systematic model in Eq.~(\ref{eq:syst}): the imaging systematics cause a positive $\Delta\Sigma_1^{(+/-)}$ signal and do not contribute to $\Delta\Sigma_2^{(+/-)}$.
For a quantitative check we can scale this plot by the average $\Sigma_{\rm crit}$ factor of our measurements, $\Sigma_{\rm crit} \sim \, 8000 \, \mhpc$.
The magnitude of the best-fit power law systematic in Fig.~\ref{fig:foreground} is $\sim 0.001 \pm 0.0005$ at $R \sim 0.1 \mpch$.
Then multiplying by $8000 \, \mhpc$ we have $8 \pm 4 \, \mhpc$ at $R \sim 0.1 \mpch$.
Thus, the fit is only $2\sigma$ away from zero, with the mean about four times the estimate from the upper-edge of the $1\sigma$ contour of Fig.~\ref{fig:epsilon}.
If we allow for negative amplitudes of the systematic, the level shifts lower to be consistent with zero at $\sim 0.0007 \pm 0.0007$.
In this case the mean systematic is between two and three times higher than the $1\sigma$ edge of Fig.~\ref{fig:epsilon}.
While this is a higher value for the systematic, note that only the first 2-3 black points are significantly positive.
The estimated systematic from this method could then be aggressively removed by discarding the first few points in the measurement.
We have tested the effect of doing so, and find that discarding the first 2 (3) points for all four estimators results in a detection significance of $3.4\sigma$ ($2.3\sigma$).
Since the detection significance with all points included is $4\sigma$, there does not appear to be any significant contamination from possible systematics at small $R$.

\subsection{Galaxy-Halo Misalignment}
\label{sec:align}

Another potential systematic that could dilute the halo axis ratio measurements is misalignment between the position angles of the LRG light and dark matter distributions.
While our upper bounds on axis ratios are robust even in the presence of misalignment, the best-fit constraints are not.
If our best-fit axis ratio of 0.78 is large compared to expectations from theory and other measurements, then misalignment is a possible cause.
\citet{acd15} found axis ratios of $q \sim 0.65$ for halos with mass $\sim 10^{11.6} M_\odot/h$, falling to $\sim 0.6$ for their largest mass bin of $\sim 5 \times 10^{12} M_\odot / h$.
Our best-fit axis ratio value is significantly larger than 0.6, but their most massive bin is still about a factor of 10 smaller than our LRG halos, so it is difficult to make a direct comparison.
In addition, \citet{oto10} measured a best-fit axis ratio of 0.54 for 18 clusters specifically chosen to be well-fit by an elliptical NFW profile.
Being very massive and specifically chosen to give good ellipticity constraints, these clusters may give a lower boundary on the expected halo axis ratios.
Given the additional uncertainty in galaxy-halo alignment distributions, and the interplay of this systematic with axis ratio constraints, we leave a careful comparison to theory for future work.
For now we only derive the effect of galaxy-halo misalignment on our measurements and compare two possible alignment distributions.

Any misalignment will lower the measured signal, thus requiring a smaller best-fit value of $\epsilon$ for the model to compensate.
If the angle between the galaxy and halo major axes is $\theta_{\rm off}$, then Eqs.~(\ref{eq:gplus}, \ref{eq:gcross}) become
\bea
\Sigma_{\rm crit} \, \gamma_+ & = & (\epsilon/2) \, [\Sigma_0 \eta_0 - I_1 - I_2] \, \cos{2(\theta-\theta_{\rm off})} \\
\Sigma_{\rm crit} \, \gamma_\times & = & (\epsilon/2) \, [-I_1 + I_2] \, \sin{2(\theta-\theta_{\rm off})} \, .
\eea
Expanding the $\cos{2(\theta-\theta_{\rm off})}$ and $\sin{2(\theta-\theta_{\rm off})}$ factors yields terms proportional to $\cos{2\theta_{\rm off}}$ and $\sin{2\theta_{\rm off}}$.
Inserting into Eqs.~(\ref{eq:gamma1}, \ref{eq:gamma2}) and integrating over the distribution of alignments $P_{\rm align} (\theta_{\rm off})$, the $\sin{2\theta_{\rm off}}$ terms average to zero.
(Halos are equally likely to be misaligned on either side of the galaxy major axis, so that $P_{\rm align}(\theta_{\rm off}) = P_{\rm align}(-\theta_{\rm off})$ by symmetry.)
This leaves
\bea
\langle \Sigma_{\rm crit} \gamma_1 \rangle & = & \hspace{-0.5em} D (\epsilon/4) \, [(2 I_1 -\Sigma_0 \eta_0) \cos{4\theta} + 2 I_2 - \Sigma_0 \eta_0] \\
\langle \Sigma_{\rm crit} \gamma_2 \rangle & = & D (\epsilon/4) \, [2 I_1 - \Sigma_0 \eta_0] \sin{4\theta} \, ,
\eea
where the dilution factor due to misalignment is given by
\be
D = \int_{-\pi/2}^{\pi/2} \de \theta_{\rm off} P_{\rm align} (\theta_{\rm off}) \cos{2\theta_{\rm off}} \, .
\ee
Thus misalignment is degenerate with halo ellipticity.

In order to gauge the range of possibilities for misalignment, we compare various models to see how each would dilute the signal.
The measurements of the large scale correlations of SDSS LRG's by \citet{ojl09} show a signal over a wide range of scales (larger than those investigated here).
By comparison with CDM dark matter halos in N-body simulations, the authors interpret the results as consistent with a misalignment of the LRG light with its parent halo, modeled as a Gaussian with a dispersion of $\sigma = 35^\circ$:
\be
P_{\rm align} (\theta_{\rm off}) \propto e^{- \theta_{\rm off}^2 / 2 \sigma^2} \, .
\ee
The measurements of \citet{ljf13} show that this broad Gaussian distribution also helps to explain the alignment of CMASS galaxies with large-scale-structure.
Assuming this misalignment distribution, we find a large dilution factor $D \approx 0.50$.
Applying this misalignment distribution of \citet{ojl09} to our measurement, we obtain a best-fit axis ratio of $\sim 0.6$, consistent with the simulations of \citet{acd15}.

However, a contrasting result is given by \citet{tmd14}, which used hydrodynamical simulations to study the alignment of dark matter halos with their central galaxies.
For their largest mass bin with $M_{\rm halo} > 10^{13} M_\odot / h$ at $z = 0.3$, the sample closest to our LRGs, \citet{tmd14} found the 2D alignment distribution
\be
P_{\rm align} (\theta_{\rm off}) \propto \, e^{-9.28 \, |\theta_{\rm off}|}
\ee
accurately describes their simulation results.
This distribution gives a much less drastic misalignment dilution, $D \approx 0.95$.
Note however, that the results between different hydrodynamical codes have not converged: details of the subgrid physics can still affect results substantially, including results for the alignment of galaxies and halos \citep{vcs15}.
Nonetheless, if the above distributions span the range of possibilities, our best fit axis ratio would fall somewhere between $0.58 - 0.77$ (ellipticity between $\epsilon = 0.25 - 0.48$).

In the presence of this theoretical uncertainty, another approach is to attempt three-point lensing methods such as the estimator of \citet{acd15}, which should produce constraints on ellipticity that are immune to galaxy-halo misalignment.
(See also \citet{spk12}.)
However, their method requires much more data than is currently available: the authors find that LSST and Euclid should provide strong detections, but ongoing surveys such as PanSTARRS, DES, and HSC may have to wait until completion before having a chance at a detection of halo ellipticity with this method.

\subsection{Miscelleneous Tests}
\label{sec:misc}

We have attempted a few more measurement tests, which we summarize briefly here.
First, as noted in \S~\ref{sec:halo-model}, Eqs.~(\ref{eq:gamma1}, \ref{eq:gamma2}) suggest estimators that involve weighting $\Delta\Sigma_2$ by $\sin{(4\theta)}$ and $\Delta\Sigma_1$ by $\cos{(4\theta)}$.
We attempted to measure with such estimators but found no improvement in the results.
Not surprisingly, constraints from $\Delta\Sigma_1$ were degraded due to averaging out the terms without $\theta$ dependence in Eq.~(\ref{eq:gamma1}).
There are no such terms in Eq.~(\ref{eq:gamma2}); however the ellipticity constraints from $\Delta\Sigma_2$ shifted downward by $1\sigma$ compared to our fiducial constraints.
The choice of weights and estimators merits further study.
Second, our $\dsom$ and $\dsop$ estimators will have some contamination from cosmic shear, which aligns nearby sources (since $\xi_+ > 0$).
This means cosmic shear acts with opposite sign to our elliptical halo lensing and with the same sign as the systematic shear that we have treated in detail in Secs.~\ref{sec:syst} and \ref{sec:syst2}.
Since cosmic shear also has roughly the $1/R$ dependence of our systematic model (see for example Fig.~8 of \citet{lin12}), to some extent it is already folded into our systematics treatment.
Third, the lens galaxies for which we have shape measurements generally have very small shear measurement errors and a significantly non-zero ellipticity magnitude.
We tried cutting on both these quantities, removing the few lenses that are poorly measured or very round.
However, neither cut affected the results significantly, likely since only a small fraction of lenses were removed by these cuts. This issue may be more important for the full galaxy population, which includes much fainter lens galaxies. 
Finally, we tested the impact of satellite LRGs since they may have poorer alignment of light with  halo mass. We  found that removing the $\sim$10 percent of LRGs in multiple systems did not change our results (the best fit ellipticity value shifted upwards by about half-sigma).

\section{Conclusions}
\label{sec:conclusion}

Using a large sample of SDSS LRGs, we have studied the quadrupole shear of the dark matter halos and ruled out the hypothesis of no ellipticity at the $4\sigma$ confidence level, and ellipticity < 0.12 at the $2\sigma$ level.
Our observable uses carefully chosen angular averages of the two components of Cartesian shear, none of which have a contribution from the usual circularly averaged monopole term.
Our estimator and modeling have advantages over previous results with tangential- and cross-shear estimators.
One component of our Cartesian estimator is insensitive to the primary systematic in the literature \citep{mhb06}, a spurious alignment of lens and source ellipticities, allowing us to isolate the shear from the halo quadrupole.
We obtain unbiased measurements from the component that is affected by the systematic by modeling and marginalizing over its effect, eliminating the need for subtracting it off and hence increasing noise.

Our key result on the halo ellipticity of LRG's is presented in Fig.~\ref{fig:epsilon}. The central value of the LRG halo ellipitcity is $0.24 \pm 0.06$ corresponding to an axis ratio of 0.78. We find that our measurements agree with simulation studies of CDM halos provided one folds in a level of misalignment of mass and light.
A Gaussian distribution of misalignment angles with $\sigma=35$ degrees, suggested by \citet{ojl09}, fits our results (though is higher than the recent simulation-based results of \citet{tmd14}).
We also measure the ellipticy of redMaPPer clusters and find results consistent with those of LRG halos, though with slightly smaller statistical significance owing to the smaller sample of suitable clusters. 

Our modeling could be further improved by simultaneously fitting for halo mass and ellipticity and then marginalizing over halo mass.
The current model has some uncertainty since we use a fixed mass and the predicted quadrupole signal is proportional to the surface density $\Sigma_0$, which is larger for more massive halos.
The $1\sigma$ mass uncertainty in \citet{msh08} for the LRG sample is $\sim 10\%$.
Consider the case where the true halo mass is 10\% higher than our assumed mass of $4\times 10^{13} M_\odot/h$.
In that case we find a true best-fit ellipticity $\epsilon = 0.22$: thus under these assumptions our fiducial best-fit value of $\epsilon = 0.24$ is an overestimate of about 8\%.

While marginalizing over mass would make us less susceptible to such biases, it would likely degrade our constraints somewhat relative to our current model with a single fixed mass and concentration.
This is because the same background sources would be used to determine both the monopole and quadrupole parameters, and the shape noise would then be slightly correlated between quadrupole and monopole elements of our combined data vector.
However, since the monopole signal is nulled with our quadrupole estimator, the signal itself would add no additional covariance.
The dependence of the signal on concentration is much weaker, and so marginalization over it is likely to be less important.
Nonetheless, concentration may be worth marginalizing over along with higher order parameters such as mass-concentration scatter and mass-luminosity scatter in future, higher signal-to-noise measurements.

Ongoing surveys (DES, KiDS, HSC) have superior image quality and depth compared to SDSS. With the potential for significantly smaller errors on halo ellipticity measurements, it will also be necessary to explore additional tests for  systematics and improve the modeling  compared to the analysis from SDSS presented here. In the previous section we have summarized a set of tests that can be pursued in greater detail. Other 
extensions could include separating  central lens galaxies; using other optical estimators of the alignment of the halo such as the distribution of neighboring galaxies; exploring other estimators for the halo ellipticity; and extending the measurements to field galaxies to study the trends with luminosity and galaxy type.
Also in the future, higher-order lensing effects such as flexion \citep{hb09} will help to increase the signal-to-noise of halo ellipticity measurements.

\subsection{Comparison to other work}

In \S~\ref{sec:mandelbaum} we compared our results in some depth with those of \citet{mhb06}.
Now we briefly survey the other attempts to measure halo ellipticity from data.
\citet{phh07} used a sample of $\sim 2\times 10^5$ lenses and $\sim 1.3\times10^6$ sources in 22 sq. deg. of single-band CFHTLS data to attempt a halo ellipticity measurement.
Without photometric redshifts the split between lenses and sources was done based simply on galaxy magnitude in the one available band.
They found a $2\sigma$ hint of elliptical halos by looking at the ratio of tangential shear along the major and minor axes of the lenses.
With data from the second Red-sequence Cluster Survey (RCS2) \citet{vhs12} used an estimator involving the ratio of tangential shear along the lens' major and minor axes, as well as the tangential- and cross-shear estimators Eqs.~(\ref{eq:cos}, \ref{eq:sin}), and none gave a significant detection.
(See also \citet{hyg04} for an earlier study of RCS.)
There has been more success in the cluster lensing regime, where the intrinsic signal of elliptical halos is higher.
Using 4,300 cluster lenses in SDSS \citet{eb09} obtained a $2.7\sigma$ detection of signal.
Furthermore, using a subset of 18 clusters that are well fit individually by an elliptical halo model, \citet{oto10} estimated a $7\sigma$ detection of ellipticity for those specific clusters.

\begin{figure}
\centering
\resizebox{85mm}{!}{\includegraphics{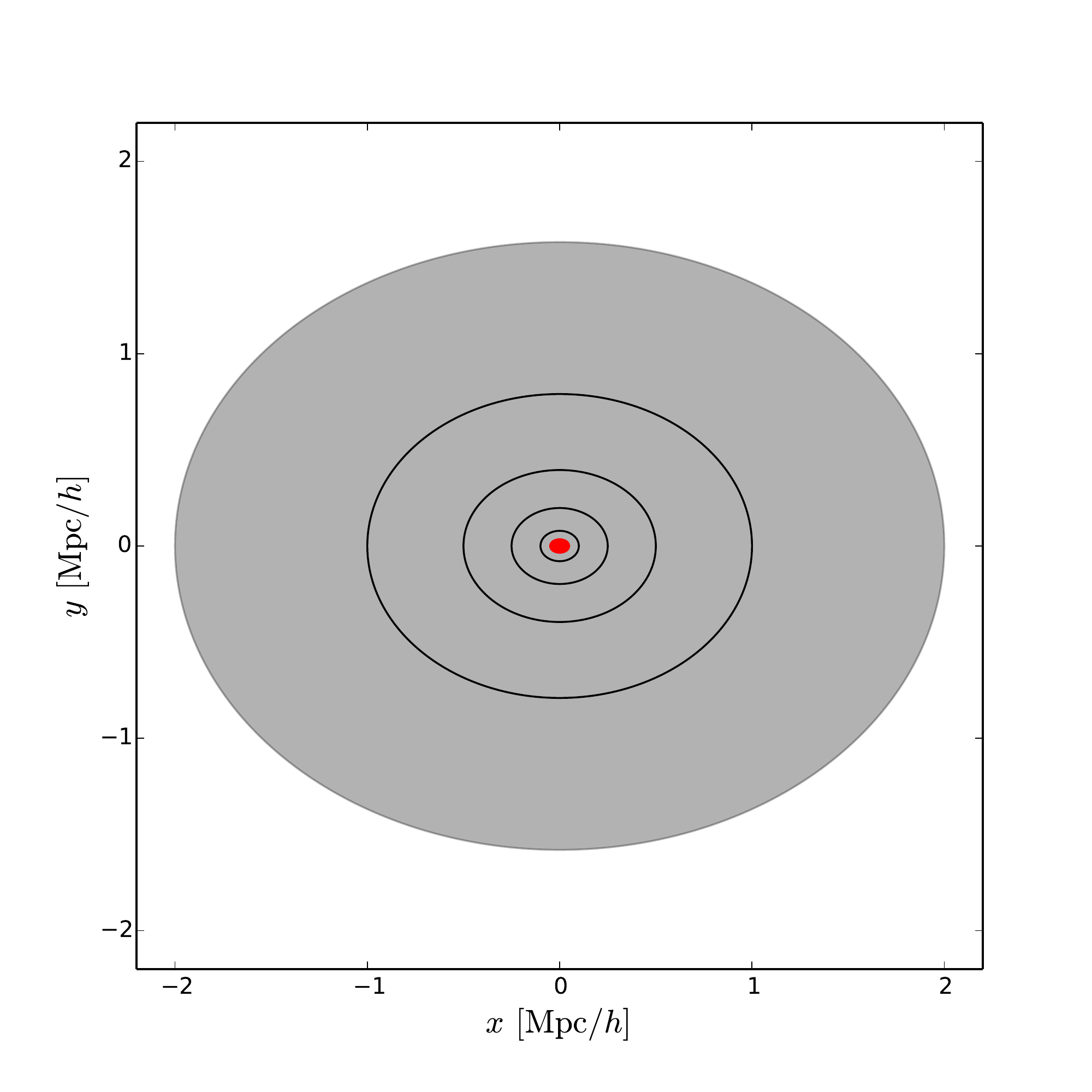}}
\caption{The ellipticity of the dark matter distribution (black ellipses) is shown over the range of scales sensitive to our measurement of quadrupole lensing.
The visible baryons (red ellipse at center) are essentially confined to scales $\sim 40$ times smaller.
Regardless of the details of the theory, any attempt to modify gravity and use just the baryons to describe lensing will produce rounder contours with increasing radius.
}
\label{fig:ellipse}
\end{figure}

\subsection{Dark matter and gravity theories}

As we have seen, galaxy-galaxy lensing provides information on halo shapes over a wide range of distances. It turns out that this range is very useful for constraining some theories of dark matter and gravity. Typically the virial radii of these galaxies are at least ten times larger than their visible sizes. Lensing measurements of the kind we have presented span this range of distances, in the case of our LRG sample it covers the range from $0.1 \mpch$ to $1 \mpch$ and beyond with the best signal-to-noise at about $0.5 \mpch$. We discuss next two examples of theories that can be tested by our measurements and future improvements. 

The lower bounds on halo ellipticities can constrain theories that attempt to do away with particle dark matter by modifying gravity.
In Fig.~\ref{fig:ellipse} we show the ellipticity of the dark matter distribution from $0.1 \mpch$ to $2 \mpch$, the range of scales sensitive to our measurement of quadrupole lensing.
The visible baryons, also pictured, are essentially confined within $0.05 \mpch$. Regardless of the details of the theory, any attempt to modify gravity and use just the baryons to describe lensing will produce rounder contours with increasing radius. Hence for MOND-like gravity theories, including recent extensions like \citet{k15}, our measurements pose a challenge. In such theories, the external tidal field can produce a quadrupole dependence on large scales, but it would be very difficult to obtain the right scale dependence over the range of our measurements.  On the other hand, halos in GR-based CDM cosmologies have a roughly scale-independent ellipticity that is consistent with our measurements. 

Another theoretical scenario of interest is self-interacting dark matter. If the dark matter particle has interactions that lead to scatterings with order unity probability over the age of the halo, then these scatterings make the halo rounder. Since the density is higher in the central regions, the effect is stronger closer to the center of the halo \citep{prb13}. Thus there is a  scale dependence that distinguishes self-interacting dark matter halos from standard CDM ones. To the extent that our ellipticity measurements match CDM predictions, we can constrain the interaction cross-section for self-interacting scenarios. Figure 3 of \citet{prb13} shows the variation of the axis ratio over $0.1$ to $1 \mpch$, a range well probed by our measurements. See also the recent study of \citet{kkl14} which examines how baryons, which may dominate in the inner parts of the galaxy, can counter the isotropizing effect of dark matter self-interactions. 

We leave the connection of our measurements to these theoretical scenarios for future work since it requires a carefully study of the scale dependence of halo ellipticity. It may be that we will need larger datasets with better lensing signal as expected from ongoing surveys to obtain meaningful constraints.

\section*{Acknowledgments}
We would like to thank the referee for a thorough reading of the paper and helpful suggestions.
We are grateful to Eric Baxter, Gary Bernstein, Neal Dalal, Yuedong Fang, Daniel Gruen, Mike Jarvis, Rachel Mandelbaum, Alan Meert, Eduardo Rozo, Eli Rykoff, Erin Sheldon, and Michael Troxel for helpful discussions. We would also like to thank Erin Sheldon for the use of his SDSS shear catalogs. BJ and JC are partially supported by the US Department of Energy grant de-sc0007901.
We benefitted from the hospitality of the Aspen Center for Physics, supported in part by National Science Foundation Grant No. PHYS-1066293.



\end{document}